# Flexible Length Polar Codes through Graph Based Augmentation

A. Elkelesh*, M. Ebada*, S. Cammerer*, S. ten Brink*
*Institute of Telecommunications, Pfaffenwaldring 47, University of Stuttgart, 70569 Stuttgart, Germany
Email: {elkelesh,ebada,cammerer,tenbrink}@inue.uni-stuttgart.de

*Abstract*—The structure of polar codes inherently requires block lengths to be powers of two. In this paper, we investigate how different block lengths can be realized by coupling of several short-length polar codes. For this, we first analyze "code augmentation" to better protect the semipolarized channels, improving the BER performance under belief propagation decoding. Several serial and parallel augmentation schemes are discussed. A coding gain of $0.3$ dB at a BER of $10^{-5}$ can be observed for the same total rate and length. Further, we extend this approach towards coupling of several "sub-polar codes", leading to a reduced computational complexity and enabling the construction of flexible length polar codes.

## I. INTRODUCTION

Polar codes were introduced by E. Arıkan [1] who showed that polar codes provably achieve capacity of any symmetric BI-DMC under successive cancellation (SC) decoding. A lot of attention has been given to polar codes due to their excellent decoding performance [2] and the fact that the code structure is explicitly given for arbitrary code rates. However, for finite length codes, the performance significantly depends on the decoding algorithm used, i.e., for the best BER performance [2] the computationally rather complex successive cancellation list (SCL) decoding [3] needs to be applied. As an alternative, a belief propagation (BP) decoder exists [4] which offers more potential for parallelization [5] and thus, high decoding throughputs at low decoding latency. Additionally, BP decoding allows efficient iterative soft-in/soft-out decoding, which may be useful, i.e., for iterative detection and decoding schemes. However, the BER performance under BP decoding is close to that under SC decoding [1] and thus, its performance has a considerable performance gap when compared to SCL decoding.

The main reason for the weaker performance of finite length polar codes, when compared to infinite length polar codes, is owed to the fact that the synthesized bit channels are not fully polarized [1]. In previous work [6], and later in [7], it has been shown that the performance of BP decoding can be improved by augmentation using an LDPC code, such that the semipolarized bit channels are further protected. In this paper, rather than using an LDPC code, an auxiliary *polar* code is applied instead over the semipolarized bit positions. In fact, the appended polar code is not an outer code in the sense that an outer code would handle all of the information bits prior to the "inner" polar code; rather, one would consider it as a short *auxiliary* code that improves protection only of some of the semipolarized channels. The approach of using a polar code (rather than an LDPC code) for code augmentation appears to be natural and elegant for hardware implementation, as the whole decoding circuit is based on the same processing elements (PE). As will be shown later, the proposed setup is able to outperform the reported gains of [6] with a comparable computational complexity. As it turns out, this idea can be extended to a parallel coupling structure by appropriate interleaving, facilitating the construction of longer polar codes based on short "sub-polar codes".

As the basic concept of polar codes can be described by the concept of channel combining and channel splitting of two bit channels, a polar codeword inherently is constrained to block lengths of $N = 2^n$ [1]. Thus, the flexibility of the block length is quite limited[1], which can be seen as one drawback whenever it comes to standardization and compatibility with existing solutions. As, typically, in many modern communication standards the block length is not a power of two, we investigate the possibility of constructing polar codes with a flexible length, such as, e.g., $N = 1536$ bits and $N = 3072$ bits, through parallel coupling of different length polar codes. Although puncturing of polar codes [9] and shortening [10] is possible, we show an alternative way and study the performance of this new setup with respect to the original construction of polar codes [1].

The paper is organized as follows: Section II briefly reviews the basic concepts of polar codes. Section III outlines how polar codes can be serially augmented to better protect the semipolarized bit channels. Section IV is devoted to constructing flexible length polar codes via parallel augmentation, facilitating flexible length codes and at potentially reduced overall computational complexity. Finally, Section V renders some conclusions.

## II. POLAR CODES

For a quick review, we briefly discuss channel polarization, code construction, as well as polar encoding and its decoding concepts [1].

### A. Channel Polarization

Channel polarization is the concept upon which polar codes are based, in which $N$ distinct channels $\left\{W_N^{(i)} : 1 \leq i \leq N\right\}$ are synthesized, starting from $N$ independent copies of a BI-DMC. The $N$ synthesized channels are polarized and have channel symmetric capacity either close to 0 (i.e., "noisy channels") or close to 1 (i.e., "noiseless channels"). These channels become perfectly noisy/noiseless as $N$ approaches infinity [1]. The process of channel polarization consists mainly of two phases:

1) **Channel Combining:** where $N$ distinct channels are created in $n = \log_2(N)$ steps, through recursively combining $N$ copies of a BI-DMC to form a vector channel $W_N : X^N \to Y^N$, where $N$ is constrained to be a power of two, i.e., $N = 2^n, n \geq 0$.

2) **Channel Splitting:** where the channel $W_N$ is split into $N$ binary-input channels $W_N^{(i)} : X \to Y^N \times X^{i-1}, 1 \leq i \leq N$. For any two channels resulting from channel combining and splitting, their Bhattacharyya parameters $Z(W)$ can be related as follows:

$$Z(W^-) \leq 2Z(W) - Z(W)^2 \quad (1)$$
$$Z(W^+) = Z(W)^2 \quad (2)$$

where $Z(W) \neq 0$ being a reliability measure [1].

---

[1]Remark: A replaced kernel function [8] also enables different length constructions; however, this is an open field of research and its flexibility is limited to the basic kernel used.

### B. Code Construction

The code construction phase is the process of computing the set of indices of the bit channels $\mathbb{A}$ on which data is transmitted. After $\log_2 N$ stages of channel polarization, $N$ distinct copies $\{W_N^{(i)}\}$ of the original channel $W$ are synthesized, each with its own $Z(W_N^{(i)})$ according to the relations emphasized in (1) and (2). Intuitively, the following property holds: $Z(W_N^{(i)}) \leq Z(W_N^{(j)})$ for all $i \in \mathbb{A}, j \in \bar{\mathbb{A}}$, and $\bar{\mathbb{A}} \cup \mathbb{A} = \{1, 2, \ldots, N\}$ [1]. There is no such code construction algorithm that computes the $Z(W_N^{(i)})$-parameters in an efficient manner for a general channel. Several research work has been conducted pursuing an efficient code construction algorithm [1][11][12].

### C. Polar Encoding

A polar code of length $N = 2^n$ is encoded using the polar code generator matrix $\mathbf{G}$ of size $N \times N$. Thus, a block of length $N$ consisting of frozen and non-frozen bits is multiplied by $\mathbf{G}$ to produce the polar codeword. The $\mathbf{G}$-matrix is $\mathbf{G} = \mathbf{F}^{\otimes n}$, where $\mathbf{F}^{\otimes n}$ denotes the $n^{th}$ Kronecker power of $\mathbf{F} = \begin{bmatrix} 1 & 0 \\ 1 & 1 \end{bmatrix}$, which is Arıkan's proposed kernel [1].

### D. Belief Propagation (BP) Decoding of Polar Codes

The design of a reliable polar decoder with the lowest possible complexity has been an active area of research, among other aspects related to polar coding. The trade-off between the coding gain and hardware complexity plays an important role in comparing different polar decoders. For instance, the SCL polar decoder enjoys larger coding gain than the BP decoder [2]. However, it suffers from higher decoding complexity $\mathcal{O}(L \cdot N \log_2 N)$, where $L$ is the list size.

The BP decoder is based on a message passing algorithm that decodes the received channel output and computes its estimates through iterations, according to a specific version of a factor graph which corresponds to a particular encoder structure. Both SC and BP decoders undergo the decoding process based on the same generator matrix. However, they differ in the following aspects:

1) **The decoding schedule:** Unlike the recursive sequential schedule of SC decoding, all the decoding nodes (PEs) are activated in one BP iteration.
2) **No intermediate hard decisions:** Unlike the serial nature of the SC decoder where each decoding stage takes into account the values of the previously hard-decoded bits, a BP decoder works iteratively where no hard decision is taken until the limit of the number of iterations is reached.
3) **No error propagation:** In contrast to the SC decoder, a BP decoder encounters no error propagation from the previously hard-decoded bits, which is attributed to be one of the main reasons why it outperforms a corresponding SC decoder in terms of BER performance.
4) **Potential for parallelization:** BP decoding shows much more potential for parallelization and thus offers more flexible implementation options than SC.

A factor graph of $N = 4$ and $N = 8$ polar codes is shown in Fig. 3a. The factor graph consists of $n = \log_2(N)$ stages, each stage consisting of $N$ nodes. There are two types of Log-Likelihood Ratio (LLR) messages: the right-to-left messages (**L**-messages) and the left-to-right messages (**R**-messages). One BP iteration consists of two update propagations:

1) Right-to-left propagation: the **L**-messages are updated starting from the rightmost stage (i.e., the stage of channel information) until reaching the leftmost stage.
2) Left-to-right propagation: the **R**-messages are updated starting from the leftmost stage (i.e., the stage of a priori information) until reaching the rightmost stage.

The output from each two nodes is the input to a specific neighboring PE, shown in [13, Fig. 1]. One PE updates the **L**- and **R**-messages as follows:

$$\begin{array}{ll} L_{out,1} = & f(L_{in,1}, L_{in,2} + R_{in,2}) \\ R_{out,1} = & f(R_{in,1}, L_{in,2} + R_{in,2}) \\ L_{out,2} = & f(R_{in,1}, L_{in,1}) + L_{in,2} \\ R_{out,2} = & f(R_{in,1}, L_{in,1}) + R_{in,2} \end{array} \quad (3)$$

where $f(a,b) = \ln\left(\frac{1+e^{a+b}}{e^a+e^b}\right)$ is commonly referred to as "box-plus" operator [13].

Finally, when the limit on the number of iterations is reached, the decoded bits at the leftmost stage are given by the estimated information vector $\hat{\mathbf{u}}$, whereas the decoded bits at the rightmost stage represent the estimated codeword $\hat{\mathbf{x}}$. The final hard decision is taken on the respective LLRs computed according to

$$\begin{array}{ll} L(\hat{u}_i) = & L_{1,i} + R_{1,i} \\ L(\hat{x}_i) = & L_{n+1,i} + R_{n+1,i} \end{array} \quad (4)$$

where $L(\hat{u}_i)$ and $L(\hat{x}_i)$ are the LLRs of the estimated *message* and the estimated transmitted codeword, respectively [5], [14].

## III. APPLYING AN AUXILIARY POLAR CODE OVER SEMIPOLARIZED CHANNELS

For finite length polar codes, a portion of the synthesized channels are semipolarized, thus bit errors over these channels are inevitable. This fact is used in [6] and [7] in which a short auxiliary LDPC code is used to protect the bits transmitted over the semipolarized channels, leading to an improved BER performance. However, a more natural and elegant approach may be to use an auxiliary polar code to protect the semipolarized bit channels of the inner polar code, as depicted in Fig. 1, which we refer to as "serial" augmentation. The use of an auxiliary polar code instead of an auxiliary LDPC code appears to be more natural as the whole system has a deterministic structure (except for the pseudo-random interleaver) with a low encoding and decoding complexity. Changing the total rate of the whole system can be done by changing the number of non-frozen bit channels in the auxiliary short polar code or in the inner polar code, which appears to be easier and more flexible than changing the rate in the proposed setups of [6] and [7].

The encoding structure of the proposed system is shown in Fig. 1. The inputs to the first (short) *auxiliary* polar encoder are the information bits ($K_1$ bits) and the frozen bits ($F_1$ bits), the output is the codeword ($N_1$ bits). This codeword is passed through a pseudo-random interleaver $\pi$ and then the output is considered to be the information bits that will be loaded to the semipolarized channels in the second "inner" polar encoding step. An interleaver is required between the two polar encoders in order to make the LLRs involved in iterative decoding close to statistically independent, at least during the first few decoding iterations. The inputs to the second polar encoder are the information bits that will be loaded to the good bit channels ($K_2$ bits), the frozen bits ($F_2$ bits) that will be loaded to the frozen bit channels (perfectly known) and the interleaved codeword from the first polar code to be loaded to the semipolarized bit channels. The output from the second polar encoder is the codeword of length $N$ bits that will be transmitted through the channel.

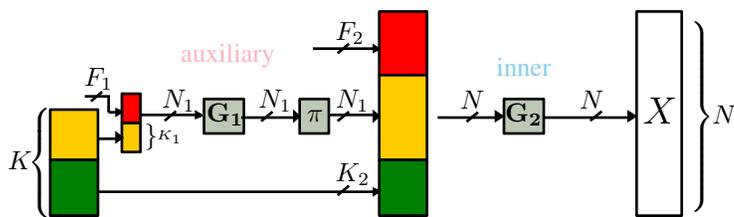

Figure 1: Proposed encoder.

The semipolarized channels, upon which the auxiliary polar code is applied, are the channels with intermediate Bhattacharyya parameter value $Z\left(W_N^{(i)}\right)$ as shown in Fig. 2 following the basic setup of [6]. Although the Bhattacharyya parameter is an error probability measure under sequential decoding [1], it turns out that, empirically, it is still a suitable measure for selecting semipolarized channels under BP decoding as proposed in [6]. For a specific set of thresholds $\delta_1$ and $\delta_2$, with $0 < \delta_1 \leq \delta_2 < 1$, three sets of channels can be defined:
1) good channels, $Z\left(W_N^{(i)}\right) \leq \delta_1$
2) intermediate channels, $\delta_1 < Z\left(W_N^{(i)}\right) \leq \delta_2$, and
3) bad channels, $Z\left(W_N^{(i)}\right) > \delta_2$.

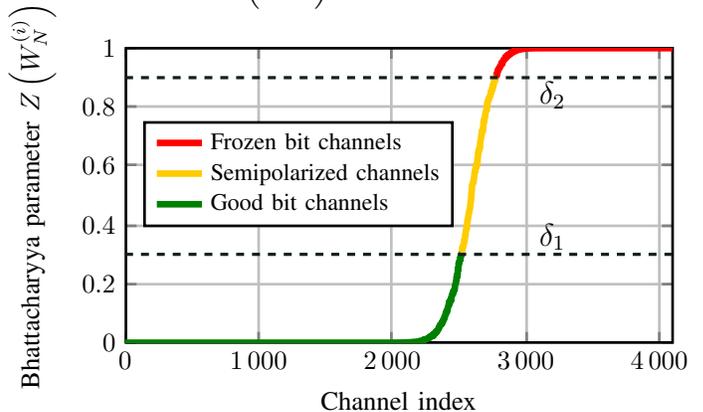

Figure 2: Sorted Bhattacharyya parameters for $N = 4096$.

The total encoding rate is given by $R_{total} = \frac{K_1+K_2}{N}$, the auxiliary polar code has a rate $R_{polar_1} = \frac{K_1}{N_1}$ and the inner polar code has a rate $R_{polar_2} = \frac{K_2+N_1}{N}$.

The corresponding decoder is shown in Fig. 3a. It is an extended version of the conventional BP decoding factor graph. The BP decoder (or the factor graph) of the auxiliary polar code is connected to the leftmost stage of the BP inner polar decoder. The decoder can be also seen in Fig. 3b in a more abstract form.

The decoding process works as follows:
1) The inner BP polar decoder receives the channel output vector $\mathbf{L}_{ch}$, then the $\mathbf{R}_2$-messages propagate from left to right, then the $\mathbf{L}_2$-messages propagate from right to left until reaching stage 1, where $\mathbf{R}_2$ and $\mathbf{L}_2$ represent the $\mathbf{R}$- and $\mathbf{L}$-messages of the inner BP decoder.
2) LLR-messages $\mathbf{L}_{2_{i,1}}$ are passed through the deinterleaver, and the output is passed to the BP polar decoder of the auxiliary code $\left(\mathbf{L}_{1_{i,n_1+1}}\right)$.
3) The BP decoder of the auxiliary polar code then performs one BP iteration (i.e., one $\mathbf{L}_1$-messages propagation and one $\mathbf{R}_1$-messages propagation), where $\mathbf{R}_1$ and $\mathbf{L}_1$ represent the $\mathbf{R}$- and $\mathbf{L}$-messages of the auxiliary BP decoder, respectively.
4) Next, the LLR-messages at the rightmost stage of the auxiliary BP decoder $\left(\mathbf{R}_{1_{i,n_1+1}}\right)$ are passed through the interleaver, and the output is passed to the BP polar decoder of the inner polar code $\left(\mathbf{R}_{2_{i,1}}\right)$. One inner code BP iteration is followed by one auxiliary code BP iteration until reaching a maximum number of iterations.
5) A hard decision is taken to estimate the message $\hat{\mathbf{u}}$ as follows
   a) The LLRs of the estimated message at the input of the auxiliary polar code
   $$L(\hat{u}_{1_i}) = L_{1_{i,1}} + R_{1_{i,1}} \qquad (5)$$

b) The LLRs of the estimated message at the input of the inner polar code

$$L(\hat{u}_{2_i}) = L_{2_{i,1}} + R_{2_{i,1}} \quad (6)$$

Recall that, the computational complexity of a polar BP decoder is proportional to the number of PEs in the respective factor graph. The polar factor graph consists of $\log_2(N)$ stages, with $N/2$ processing elements per stage. Thus, a total of $\log_2(N) \cdot \frac{N}{2}$ processing elements per factor graph.

Throughout this paper, the channel used is an AWGN channel, the modulation is BPSK, the total code rate $R = 0.5$ and polar codes are constructed based on Arıkan's Bhattacharyya bounds [1] of bit channels designed at SNR of $E_s/N_0 = 0\,\text{dB}$.

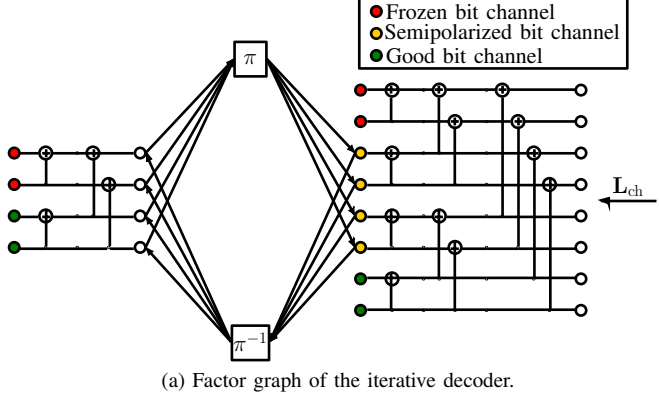

(a) Factor graph of the iterative decoder.

(b) Information flow, simplified (left) and detailed (right).

Figure 3: BP information flow of the combined inner and auxiliary polar decoder.

One setup (setup 1) is proposed in this section, with code parameters given in Tab. I and structure described in Fig. 1. A BER performance comparison between the $N = 4096$ polar code under conventional BP decoding and setup 1 is shown in Fig. 4. A $0.3\,\text{dB}$ coding gain is achieved at BER of $10^{-5}$ on the "small" expense of computational complexity as shown in Tab. II, due to the introduction of the auxiliary polar code.

## IV. COUPLING FOR FLEXIBLE LENGTH POLAR CODES

The proposed setup of the previous section can be extended to accommodate an extra inner polar code. Thus, the new setup will consist of two inner polar codes connected, or coupled, through one short auxiliary polar code, which is also used to protect the information bits transmitted on the semipolarized bit channels of the two inner polar codes; this is what we refer to as "parallel" augmentation. The semipolarized channels are chosen in the same manner as they were chosen in [6], using the Bhattacharyya parameter value $Z(W_N^{(i)})$. The encoder of this new, extended setup is shown in Fig. 5.

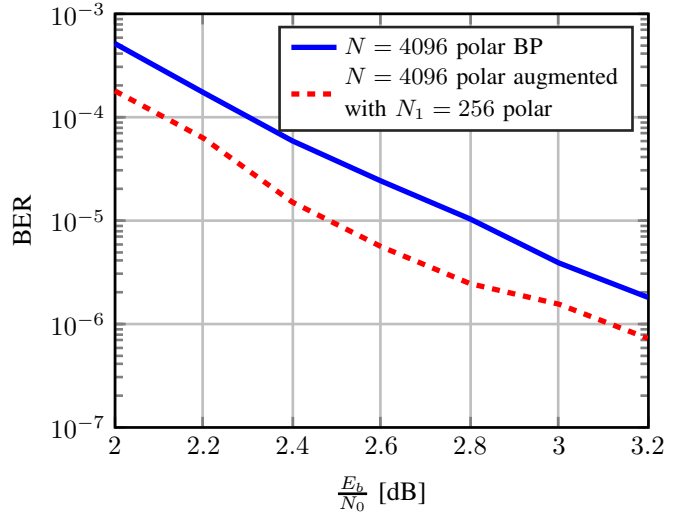

Figure 4: BER-curves of the different setups ($N = 4096$), compare to Tab. I.

Table I: Simulation parameters used for different setups.

| Setup | $K_i$ | $F_i$ | $N_i$ |
|---|---|---|---|
| 1 | $K_1 = 128$<br>$K_2 = 1920$<br>$K = \sum_i K_i = 2048$ | $F_1 = 128$<br>$F_2 = 1920$ | $N_1 = 256$<br>$N_2 = 4096$<br>$N = 4096$ |
| 2 | $K_1 = 128$<br>$K_2 = 960$<br>$K_3 = 448$<br>$K = \sum_i K_i = 1536$ | $F_1 = 128$<br>$F_2 = 960$<br>$F_3 = 448$ | $N_1 = 256$<br>$N_2 = 2048$<br>$N_3 = 1024$<br>$N = 3072$ |
| 3 | $K_{1,2,3,4} = 64$<br>$K_{5,6,7,8} = 448$<br>$K = \sum_i K_i = 2048$ | $F_{1,2,3,4} = 64$<br>$F_{5,6,7,8} = 448$ | $N_{1,2,3,4} = 128$<br>$N_{5,6,7,8} = 1024$<br>$N = 4096$ |

Table II: Number of processing elements per setup.

| Setup | Number of PEs = $\sum_i \log_2(N_i) \cdot \frac{N_i}{2}$ |
|---|---|
| $N = 4096$ polar BP | 24576 (reference) |
| 1 | 25600 |
| 2 | 17408 |
| 3 | 22272 |

Note that, the total number of transmitted bits is $N = N_2 + N_3$, given that $N_2, N_3 > N_1$, where $N_1$ is the code length of the auxiliary polar code. The total number of information bits conveyed per codeword is $K = K_1 + K_2 + K_3$. Thus the total encoding rate is $R_{total} = \dfrac{K_1 + K_2 + K_3}{N_2 + N_3}$, the auxiliary polar code has a rate $R_{polar_1} = \dfrac{K_1}{N_1}$, the first inner polar code has a rate $R_{polar_2} = \dfrac{K_2 + \frac{N_1}{2}}{N_2}$ and the second inner polar code has a rate $R_{polar_3} = \dfrac{K_3 + \frac{N_1}{2}}{N_3}$. The decoding algorithm is similar to that of the setup in the previous section, as depicted in Fig. 6.

Finally, for $N_2 \neq N_3$, implementing *flexible* length codes become possible. Tab. I shows the full code parameters for one example of the proposed scheme (setup 2). The effective code length of setup 2 is $N = 3072$ bits. Fig. 7 shows that setup 2 outperforms the conventional polar code under BP decoding in terms of BER, even with a shorter code length $N$. The effect of the auxiliary polar code on the BER is shown in Fig. 7: keeping the individual polar codes of lengths $N_2$, $N_3$ separate (green BER curve) is much worse than connecting them through the auxiliary polar code (dotted red curve), i.e., the information flow in the iterative decoder through coupling works.

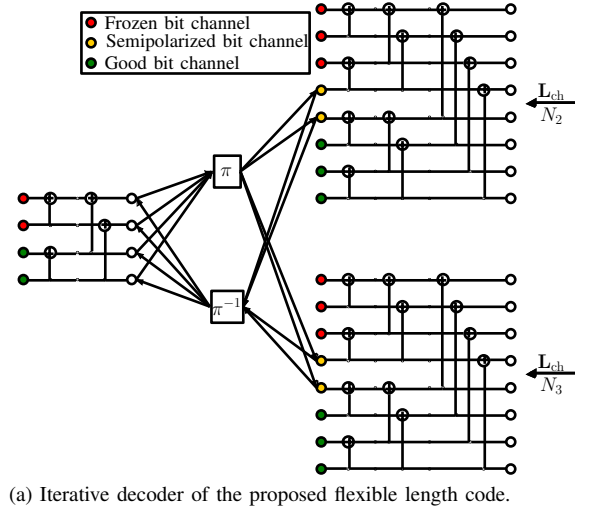

(a) Iterative decoder of the proposed flexible length code.

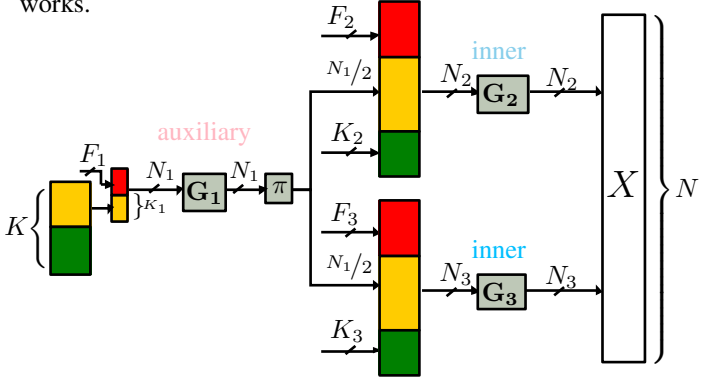

Figure 5: Encoder of the proposed flexible length code.

The proposed setup (setup 2) saves a noticeable computational complexity when compared to the conventional BP decoding of the $N = 4096$ polar code, as can be seen in Tab. II.

The above setup can be used to produce modified polar codes with flexible codeword lengths $N \neq 2^n$, mitigating the code length constraint due to the used kernel (Arıkan's $2 \times 2$ kernel $\mathbf{F} = \begin{bmatrix} 1 & 0 \\ 1 & 1 \end{bmatrix}$ [1]). This may prove to be useful when proposing polar codes, or, more precisely, "polar-like" codes for communications standards.

The same concept can be used to couple an arbitrary number of inner polar codes, which we refer to as "parallel augmentation". Thus, large codes can be built based on small polar codes, e.g., four inner polar codes ($5 \leq i \leq 8$) are coupled with four short auxiliary polar codes ($1 \leq i \leq 4$) in a ring like structure, as shown in Fig. 8. Tab. I shows the full code parameters of this setup (setup 3).

Finally, Fig. 9 shows a BER performance comparison between the $N = 4096$ polar code under conventional BP decoding and setup 3 (note that setup 3, also, has an effective codeword length of $N = 4096$). Although the BER performance of both setups is similar, setup 3 has lower overall computational complexity as shown in Tab. II. Thus, the performance of a polar code of length $N = 4096$ can be approached by four blocks of length $N = 1024$, saving in computational complexity. Again, the effect of the auxiliary polar codes on the BER is illustrated in Fig. 9 by the green (uncoupled, no auxiliary polar codes used) and red curves (coupled). Similar results and behavior is achieved in [15],

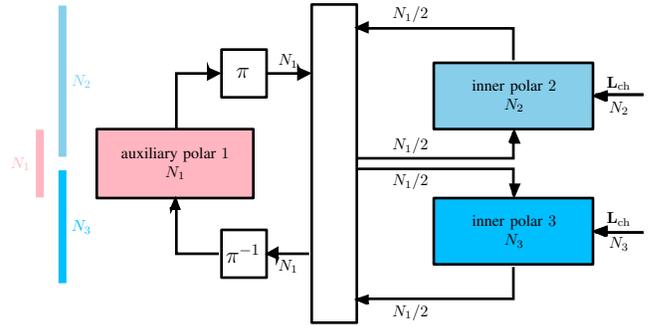

(b) Information flow, simplified (left) and detailed (right).

Figure 6: BP information flow in the decoder of the flexible length code.

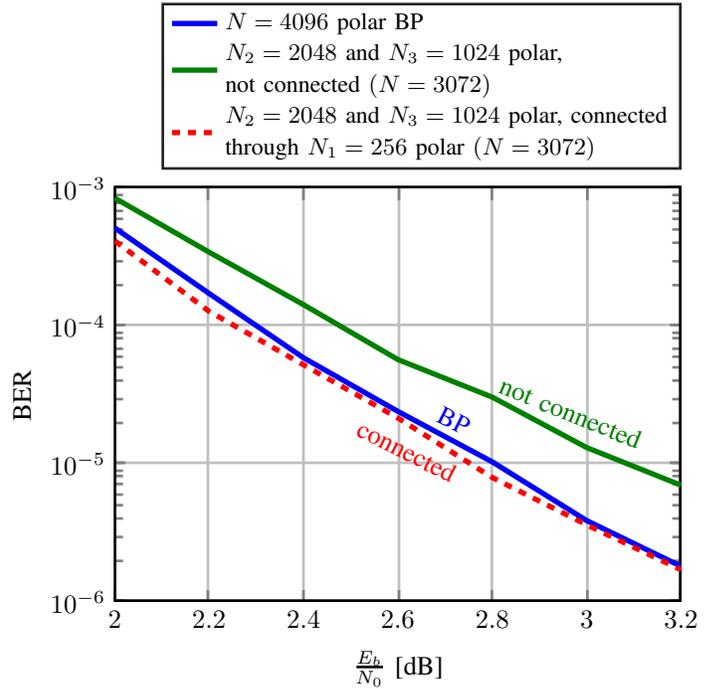

Figure 7: BER-curves of the different setups (compare to Tab. I).

where systematic polar codes are concatenated in parallel, akin to the structure of classic turbo-codes [16].

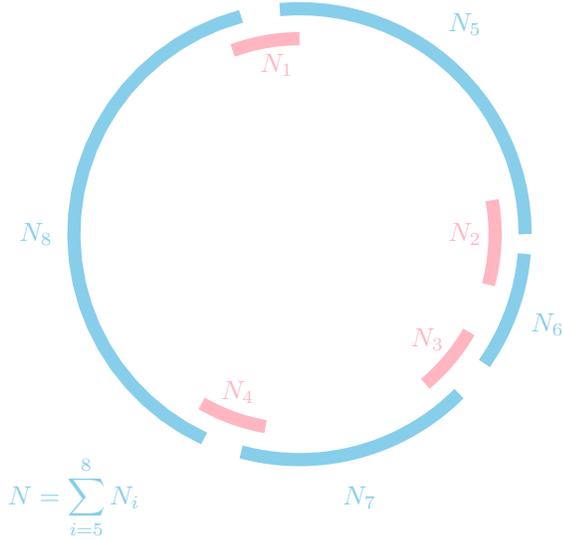

Figure 8: Four inner polar codes of lengths $N_{5,6,7,8}$ coupled, in a ring like structure, through four auxiliary polar codes of lengths $N_{1,2,3,4}$.

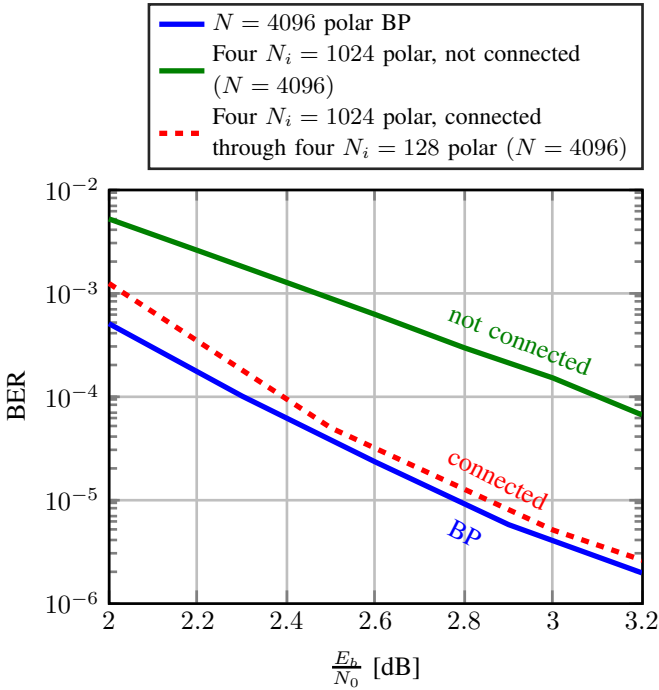

Figure 9: BER-curves of the different setups ($N = 4096$), compare to Tab. I.

## V. CONCLUSION

We introduced the idea of augmenting a finite length "inner mother" polar code by a short auxiliary (or "outer") polar code to better protect the semipolarized bit channels, leading to a coding gain of $0.3\,\mathrm{dB}$ at a BER of $10^{-5}$ under BP decoding. Moreover, several polar codes were connected through auxiliary polar codes, facilitating flexible length codes with similar BER performance at lower computational complexity when compared to the conventional BP decoder of a single, longer polar code. The successful coupling of short-length polar codes indicates that the more general concept of spatially coupling may be applicable to polar codes in the same sense as has been successfully applied to LDPC codes [17] and turbo-like codes [18] already.